\documentclass[
 reprint,
 aps,
 prl,
 superscriptaddress,
 longbibliography,
 nobibnotes,
 floatfix,
 amsmath,amssymb
]{revtex4-2}

\usepackage{graphicx}
\usepackage{dcolumn}
\usepackage{bm}
\usepackage[colorlinks=true,allcolors=blue]{hyperref}

\usepackage{braket}
\usepackage{caption}
\usepackage{subcaption}
\DeclareCaptionJustification{fulljustified}{\leftskip=0pt\rightskip=0pt\parfillskip=0pt plus 1fil\relax}
\usepackage{placeins}
\usepackage{needspace}

\begin{document}

\preprint{APS/123-QED}

\title{The radical pair mechanism can generate nuclear-spin entanglement}

\author{Rishabh}
\email{rishabh1@ucalgary.ca}
\author{Hadi Zadeh-Haghighi}
\email{hadi.zadehhaghighi@ucalgary.ca}
\author{Christoph Simon}
\email{csimo@ucalgary.ca}

\affiliation{Department of Physics and Astronomy, University of Calgary, Calgary, Alberta, Canada\\
 Institute for Quantum Science and Technology, University of Calgary, Calgary, Alberta, Canada\\
 Hotchkiss Brain Institute, University of Calgary, Calgary, Alberta, Canada}

\begin{abstract}
The radical pair mechanism is a quantum-mechanical process underlying weak magnetic field effects in chemical and biological systems. While nuclear polarization and electronic entanglement in this mechanism have been studied extensively, the possibility of creating entanglement between nuclear spins has, to the best of our knowledge, remained unexplored. Here, we show that nuclear-spin entanglement can develop from initially maximally mixed nuclear spin states and evolve over repeated radical pair cycles. Its generation and persistence depend on the initial electronic state, the available reaction channels, and the interplay between recombination rates, hyperfine interactions, and the magnetic field. We demonstrate this behaviour in both a minimal two-nucleus model and a five-nucleus model inspired by the photosynthetic radical pair P700--$A_1$. Radical pair mechanism generated nuclear-spin entanglement offers a potential new resource for quantum technologies and opens avenues for exploring its role in biological systems.

\end{abstract}

\maketitle

\begin{figure*}[t!]
    \centering

    \begin{subfigure}[t]{1\textwidth}
        \centering
        \includegraphics[width=0.8\linewidth]{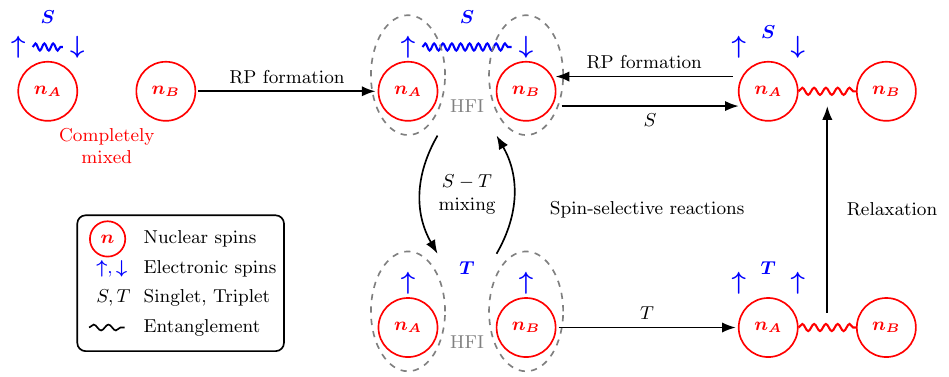}

    \caption{}
    \label{fig:1a}
    
    \hfill
    \end{subfigure}
    \begin{subfigure}[t]{0.49\textwidth}
        \centering\vspace{0pt}
        \includegraphics[width=0.9\linewidth]{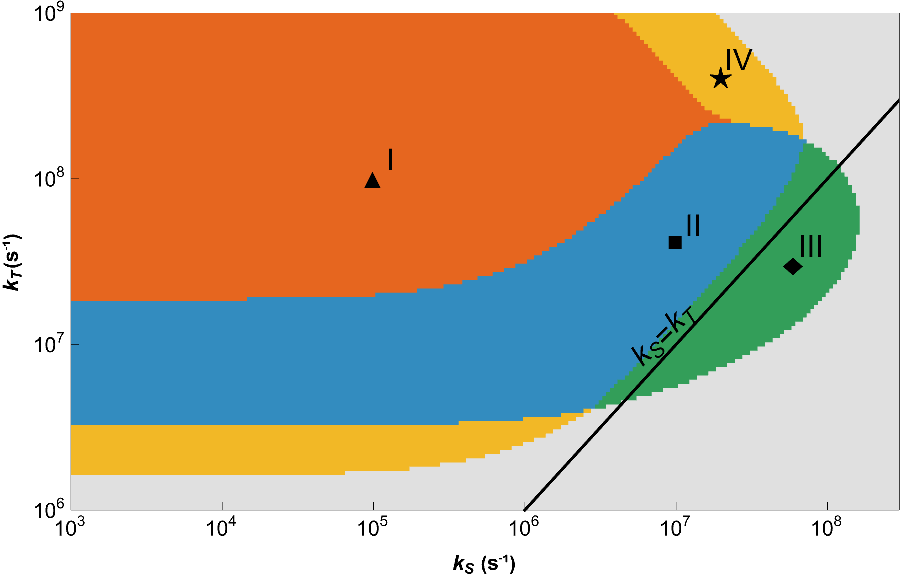}
        \caption{}
        \label{fig:1b}
    \end{subfigure}
    \begin{subfigure}[t]{0.49\textwidth}
        \centering\vspace{0pt}
        \includegraphics[width=0.9\linewidth]{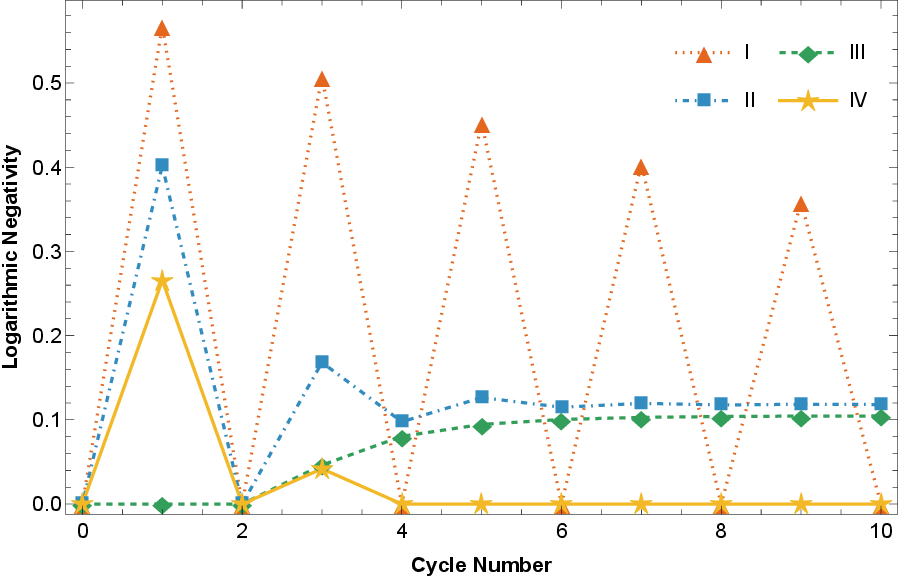}
        \caption{}
        \label{fig:1c}
    \end{subfigure}

    \caption{\textit{Nuclear-spin entanglement generation and its behaviour under repeated radical pair cycles.} (a) Schematic of repeated singlet-born radical pair (RP) cycles, with one spin-$1/2$ nucleus in each radical. Starting from maximally mixed nuclear spins, nuclear-spin entanglement is generated during RP spin dynamics in conjunction with spin-selective recombination through both singlet and triplet channels. Following recombination and electronic relaxation, the electrons are reinitialized in the singlet state. Motivated by the potentially long relaxation times of nuclear spins, we assume that their state is retained for the next cycle. (b) Classification of the ten-cycle trajectories into no entanglement detected (grey), transient entanglement (yellow), zero/nonzero oscillations (orange), nondecreasing entanglement (green), and other nonmonotonic entanglement (blue), comprising nonmonotonic behaviour that does not fall into the preceding categories. The full classification criteria are given in the Supplemental Material. The map uses a logarithmic grid with 0.025-decade spacing along both axes ($220\times121$ points), within $k_S=10^3$--$3\times10^8$ and $k_T=10^6$--$10^9~\mathrm{s}^{-1}$. Entanglement occurs predominantly where triplet recombination is faster than singlet recombination. The solid line marks $k_S=k_T$. (c) Nuclear-spin logarithmic negativity at the end of each of ten RP cycles for the representative points $\mathrm{I}$--$\mathrm{IV}$ marked in (b), with reaction-rate pairs $(k_S,k_T)$: $\mathrm{I}=(10^5,10^8)$, $\mathrm{II}=(10^7,4\times10^7)$, $\mathrm{III}=(6\times10^7,3\times10^7)$, and $\mathrm{IV}=(2\times10^7,4\times10^8)~\mathrm{s}^{-1}$. Cycle zero denotes the initial maximally mixed nuclear spin state. These trajectories illustrate the distinct behaviours mapped in (b). For (b) and (c), $a_A=a_B=1~\mathrm{mT}$, $B=0.05~\mathrm{mT}$, and $r_A=r_B=10^6~\mathrm{s}^{-1}$.}

    \label{fig:1}
\end{figure*}

\textit{Introduction---}The radical pair mechanism (RPM) is a quantum-mechanical process involving the coherent dynamics of the spin degrees of freedom of a pair of electrons and their neighbouring nuclei~\cite{closs1969mechanism,kaptein1969chemically,steiner1989magnetic}. It begins with the creation of a pair of radicals, molecules that each contain one unpaired electron, effectively forming a composite system of two spin-1/2 particles. These spins undergo transient coherent evolution driven by various interactions, particularly with neighbouring nuclear spins and external magnetic fields, leading to interconversion between states of different total electronic spin (singlet-triplet mixing). Meanwhile, radical pairs (RPs) can recombine to form chemical products. Reactions of organic radicals generally conserve electron spin angular momentum. Singlet and triplet RPs therefore recombine into products with the corresponding electron-spin multiplicity, making recombination spin selective. Altering the external magnetic field modifies the spin dynamics and thereby changes the yields of chemical products, providing a plausible basis for the influence of weak magnetic fields (WMFs) on chemical reactions involving radicals. Indeed, the RPM underlies numerous chemical and potentially biochemical reactions exhibiting sensitivity to WMFs~\cite{lawler1971,steiner1989magnetic,schulten1978biomagnetic,hore2016radical,zadeh2022magnetic,xu2021cryptochrome,ikeya2021cellular,zadeh2026tubulin,rishabh2026planarians,zadeh2026pigmentation}.

The RPM was first proposed to explain certain anomalies observed in nuclear magnetic resonance (NMR) and electron spin resonance (ESR) spectra for reactions involving radical species, phenomena now known as chemically induced dynamic nuclear polarization (CIDNP) and chemically induced dynamic electron polarization (CIDEP), respectively~\cite{closs1969mechanism,kaptein1969chemically}. For WMFs, the thermal energy at room temperature is much larger than the nuclear Zeeman splitting. Therefore, nuclear spins usually start in a completely mixed state. However, hyperfine interactions (HFIs) couple the dynamics of the electronic spins to those of the nuclear spins. This coupling can lead to nuclear spin sorting, as different nuclear spin states can favour the formation of different reaction products. Therefore, repeated RPM cycles can build up nuclear polarization~\cite{wong2021nuclear}. Nuclear spin order beyond single-spin polarization has long been recognized in CIDNP, including the multiplet effects described by Kaptein~\cite{kaptein1971rules}. More recent studies have demonstrated heteronuclear singlet order and multiple-spin order~\cite{sheberstov2021singlet,stricker2025multiple}, while nuclear coherences in reaction products and their role in polarization transfer have also been investigated~\cite{ivanov2022coherence}. These studies do not, however, establish entanglement between nuclear spins.

RPs are often created via electron transfer from a donor to an acceptor molecule. If the two electrons on the donor molecule are initially in an entangled state, such as the singlet state, spin-conserving transfer of one of these electrons to the acceptor leaves the resulting RP in the same entangled state. This fact, together with the importance of entanglement as a resource in various quantum information tasks, has led to numerous studies investigating the longevity and usefulness, if any, of electron-spin entanglement as a basis for magnetosensing based on the RPM~\cite{cai2010quantum,gauger2011sustained,kominis2012magnetic,tiersch2012decoherence,hogben2012entanglement,pauls2013quantum,zhang2014sensitivity}. More recent studies have broadened this focus to include entanglement between electron spins and neighbouring nuclei during RPM dynamics~\cite{kominis2020quantum,smart2026engineering}.

These findings raise the question of whether the RPM can also generate entanglement between nuclear spins. Here, we investigate this possibility between the nuclear-spin subsystems of the two radicals, starting from maximally mixed nuclear states, and examine how the entanglement evolves over repeated cycles.

\textit{Methods---} In order to study the evolution of nuclear-spin entanglement over multiple RPM cycles, we need to keep track of the nuclear spin state after recombination at the end of each cycle. For this purpose, we build on the approach used by Wong et al.~\cite{wong2021nuclear} to study nuclear polarization. We illustrate this approach using the example shown in Fig.~\ref{fig:1a}.

In the ground state of the donor-acceptor pair ($A$-$B$), the two electrons are paired in the singlet state in the outermost molecular orbital of $A$. The RP is formed when one of the electrons is transferred to $B$. The electron spins are preserved during the transfer process, and hence the initial electronic state of the RP is the singlet state. Each molecule contains a single spin-$\tfrac{1}{2}$ nucleus. Initially, the nuclear spins are in a completely mixed state.

The RP then undergoes coherent singlet-triplet interconversion, during which the nuclear spins become coupled to the electronic spins via HFIs, potentially generating nuclear-spin entanglement. While this spin dynamics takes place, chemical recombination occurs in parallel through spin-selective singlet and triplet reaction channels. Following triplet recombination, the donor--acceptor pair occupies an electronically excited triplet state. This electronic state subsequently relaxes to the singlet ground state while preserving the nuclear spin state. On the other hand, singlet recombination directly leads the donor–acceptor pair to the singlet ground state. An intermediate excited singlet state may occur in practice, but is not explicitly included in our model; any subsequent relaxation is assumed to preserve the nuclear spin state. Note that due to singlet--triplet mixing, a singlet-born RP can recombine through both singlet and triplet channels, provided both product states are energetically accessible~\cite{Schulten,Fay}.

The RP is then formed again, but the nuclear spins are now, in general, no longer in the completely mixed state, as their state has been modified by the preceding RPM cycle. The process is repeated for the desired number of cycles. Repeated RP formation could be driven by continued illumination or sustained electron flow, for example in an electron transport chain, provided the same nuclear-spin subsystem participates in successive cycles.

RP dynamics usually occur on nanosecond-to-microsecond timescales, whereas nuclear spin relaxation times can reach seconds~\cite{maurer2012memory}. Motivated by this difference in timescales, we neglect nuclear spin relaxation throughout the simulated sequence of cycles. Separately, we assume that recombination, electronic relaxation, and RP formation leave the nuclear spin state unchanged. In particular, unlike Wong et al.~\cite{wong2021nuclear}, we retain nuclear spin coherences between successive cycles.

Let $\hat\rho_{\mathrm n}^{(m)}$ denote the nuclear spin state at the end of cycle $m$, with $\hat\rho_{\mathrm n}^{(0)}=\hat I_{d_n}/d_n$ denoting the nuclear state before the first cycle. The first RP cycle therefore begins in
\begin{equation}
\hat\rho_{\mathrm{RP}}^{(0)}(0)=\ket S\bra S\otimes\frac{\hat I_{d_n}}{d_n}.
\end{equation}
Here $\ket S$ is the electronic singlet, $d_n$ is the dimension of the nuclear-spin Hilbert space, and $\hat I_{d_n}$ is its identity operator. For the two-nucleus model, $d_n=2\times2=4$. The superscript on $\hat\rho_{\mathrm{RP}}^{(m)}$ denotes the number of completed cycles before its formation. Its input nuclear spin state is $\hat\rho_{\mathrm n}^{(m)}$, with $m=0,1,\ldots$.

The radical pair density matrix evolves according to the Liouville master equation,
\begin{equation}
\frac{d\hat{\rho}_{\mathrm{RP}}^{(m)}(t)}{dt}
=
-\mathcal{L}\left[\hat{\rho}_{\mathrm{RP}}^{(m)}(t)\right],
\end{equation}
where the Liouvillian,
\begin{equation}
\mathcal{L}
=
i\mathcal{H}
+
\mathcal{K}
+
\mathcal{R},
\end{equation}
contains the Hamiltonian superoperator $\mathcal{H}$, the Haberkorn reaction superoperator $\mathcal{K}$, and the spin-relaxation superoperator $\mathcal{R}$.

The Hamiltonian superoperator $\mathcal{H}$ acts on the density matrix $\hat{\rho}(t)$ as
\begin{equation}
\mathcal{H}[\hat{\rho}(t)]
=
[\hat{H},\hat{\rho}(t)],
\end{equation}
where the Hamiltonian $\hat{H}$ is expressed in angular-frequency units. Only the electronic Zeeman and isotropic hyperfine interactions are included in $\hat{H}$:
\begin{equation}
\hat{H}
=
\sum_{i=A,B}
\gamma_e\hat{\mathbf{S}}_i\cdot\mathbf{B}
+
\sum_{i=A,B}
\gamma_e a_i\hat{\mathbf{S}}_i\cdot\hat{\mathbf{I}}_i,
\label{Hamiltonian}
\end{equation}
where $\gamma_e$ is the magnitude of the free-electron gyromagnetic ratio, $\hat{\mathbf{S}}_i$ and $\hat{\mathbf{I}}_i$ are the dimensionless electron and nuclear spin operators, respectively, and $a_i$ are the isotropic hyperfine coupling constants (HFCCs), expressed in mT. Rapid molecular tumbling in solution can average out anisotropic hyperfine interactions. Electron--electron exchange and dipolar couplings can be neglected when the separation between the radicals is sufficiently large.

The Haberkorn reaction superoperator~\cite{haberkorn1976density} is given by
\begin{equation}
\mathcal{K}[\hat{\rho}(t)]
=
\frac{k_S}{2}\{\hat{P}_S,\hat{\rho}(t)\} + \frac{k_T}{2}\{\hat{P}_T,\hat{\rho}(t)\},
\end{equation}
where $\{,\}$ denotes the anticommutator, $\hat{P}_{T/S}$ is the triplet/singlet projection operator and $k_{T/S}$ is the corresponding reaction rate constant.

Spin relaxation arises from the interaction of the electron spins with random, time-dependent local magnetic fields, which represent effective stochastic magnetic noise arising primarily from fluctuating spin interactions induced by the molecular motion of the radicals. The relaxation superoperator takes the form~\cite{kattnig2016electron}
\begin{equation}
\begin{aligned}
\mathcal{R}[\hat{\rho}(t)]
&=
\sum_{i=A,B} r_i
\Bigg[
\frac{3}{4}\hat{\rho}(t)
- \hat{S}_{i,x}\hat{\rho}(t)\hat{S}_{i,x}
\\ &\qquad - \hat{S}_{i,y}\hat{\rho}(t)\hat{S}_{i,y}
- \hat{S}_{i,z}\hat{\rho}(t)\hat{S}_{i,z}
\Bigg],
\end{aligned}
\end{equation}
where $r_i$ is the relaxation rate of the unpaired electron spin on radical $i$. Unless stated otherwise, both radicals experience a spin-relaxation rate of $r=10^6~\mathrm{s}^{-1}$. This is consistent with estimates of electron-spin relaxation in organic RPs, such as those in cryptochrome~\cite{kattnig2016relaxation}.

The nuclear spin state after completion of cycle $m$, $\hat\rho_{\mathrm n}^{(m)}$, is obtained by tracing over the electron spins and combining both recombination channels:
\begin{equation}
\begin{aligned}
\hat\rho_{\mathrm n}^{(m)}
={}&\frac{k_S}{2}\int_0^\infty\!\mathrm{Tr}_e
\left\{\hat P_S,\hat\rho_{\mathrm{RP}}^{(m-1)}(t)\right\}\,dt\\
&+\frac{k_T}{2}\int_0^\infty\!\mathrm{Tr}_e
\left\{\hat P_T,\hat\rho_{\mathrm{RP}}^{(m-1)}(t)\right\}\,dt.
\end{aligned}
\label{eq:cycle_output}
\end{equation}
Here $\mathrm{Tr}_e$ denotes the electron partial trace. The next RP cycle begins with the nuclear spin state produced by the preceding cycle:
\begin{equation}
\hat\rho_{\mathrm{RP}}^{(m)}(0)=\ket S\bra S\otimes\hat\rho_{\mathrm n}^{(m)}.
\end{equation}

Both recombination channels ultimately return the electrons to the same state before the next cycle. Before this reset, however, the distinct reaction pathways could affect downstream chemical processes, leaving a record of the reaction outcome outside the donor--acceptor pair. Within the pair, the nuclear spins can retain memory of the preceding dynamics and influence subsequent reaction yields. Our model accounts for this nuclear memory but does not explicitly describe downstream signalling.

The entanglement between the nuclear spins is characterized using the logarithmic negativity. Let $\lambda_i$ denote the eigenvalues of the nuclear spin density matrix after partial transposition with respect to one of the nuclear-spin subsystems, say $n_A$. The logarithmic negativity at the end of cycle $m$ ($E_N^{(m)}$) is then defined as~\cite{vidal2002computable}:
\begin{equation}
E_N^{(m)}
=
\log_{2}\Big(\sum_i |\lambda_i|\Big).
\end{equation}

\begin{figure}[t]
    \centering
    
\begin{subfigure}{\linewidth}
    \centering
    \includegraphics[width=\linewidth]{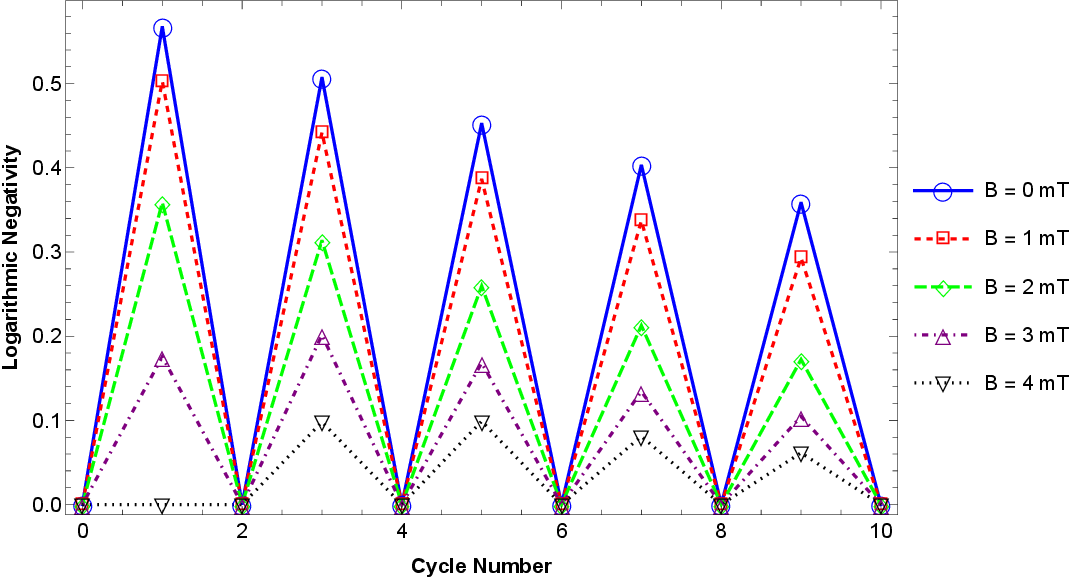}
    \caption{}\label{fig:2a}
    \end{subfigure}
    \par\medskip
    \begin{subfigure}{\linewidth}
    \centering
    \includegraphics[width=\linewidth]{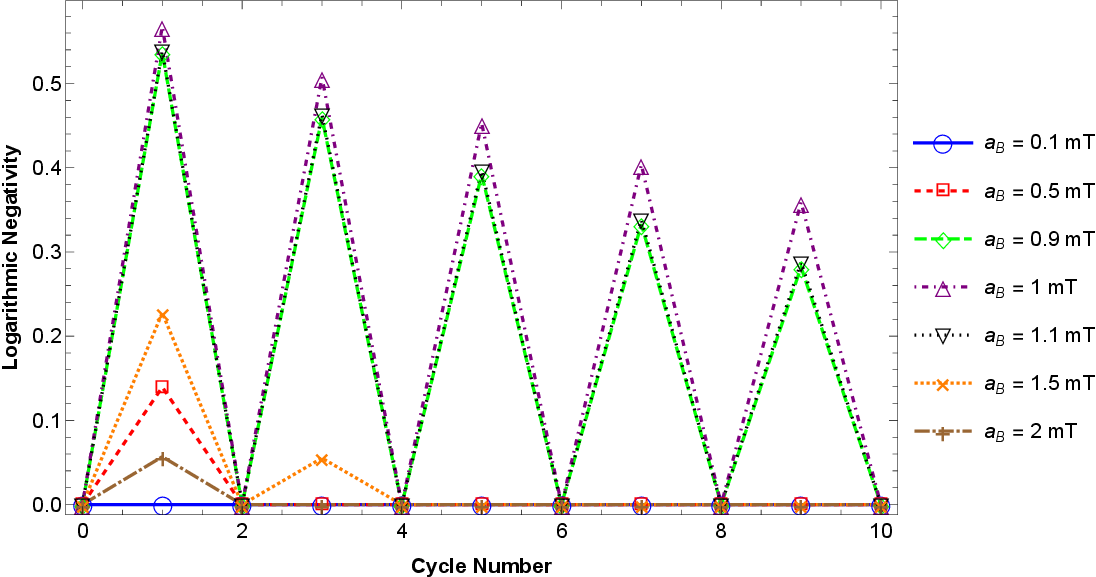}
    \caption{}\label{fig:2b}
    \end{subfigure}
    \caption{\textit{Magnetic field and hyperfine coupling dependence of nuclear-spin entanglement.} Nuclear-spin logarithmic negativity at the end of each of ten singlet-born RP cycles, starting from maximally mixed nuclear spins. (a) Magnetic field dependence for equal hyperfine couplings, $a_A=a_B=1~\mathrm{mT}$. The largest entanglement occurs at zero field, and increasing the field generally reduces the oscillation amplitudes. (b) Hyperfine coupling dependence at $B=0.05~\mathrm{mT}$, with $a_A=1~\mathrm{mT}$ fixed. Couplings $a_B$ close to $a_A$ produce larger entanglement over successive cycles. Both panels use $k_S=10^5~\mathrm{s}^{-1}$, $k_T=10^8~\mathrm{s}^{-1}$ (point $\mathrm{I}$ in Fig.~\ref{fig:1b}), and $r_A=r_B=10^6~\mathrm{s}^{-1}$. Cycle zero denotes the initial nuclear spin state.}
    \label{fig:2}
\end{figure}

\begin{figure}[ht!]
    \centering
    \includegraphics[width=1\linewidth]{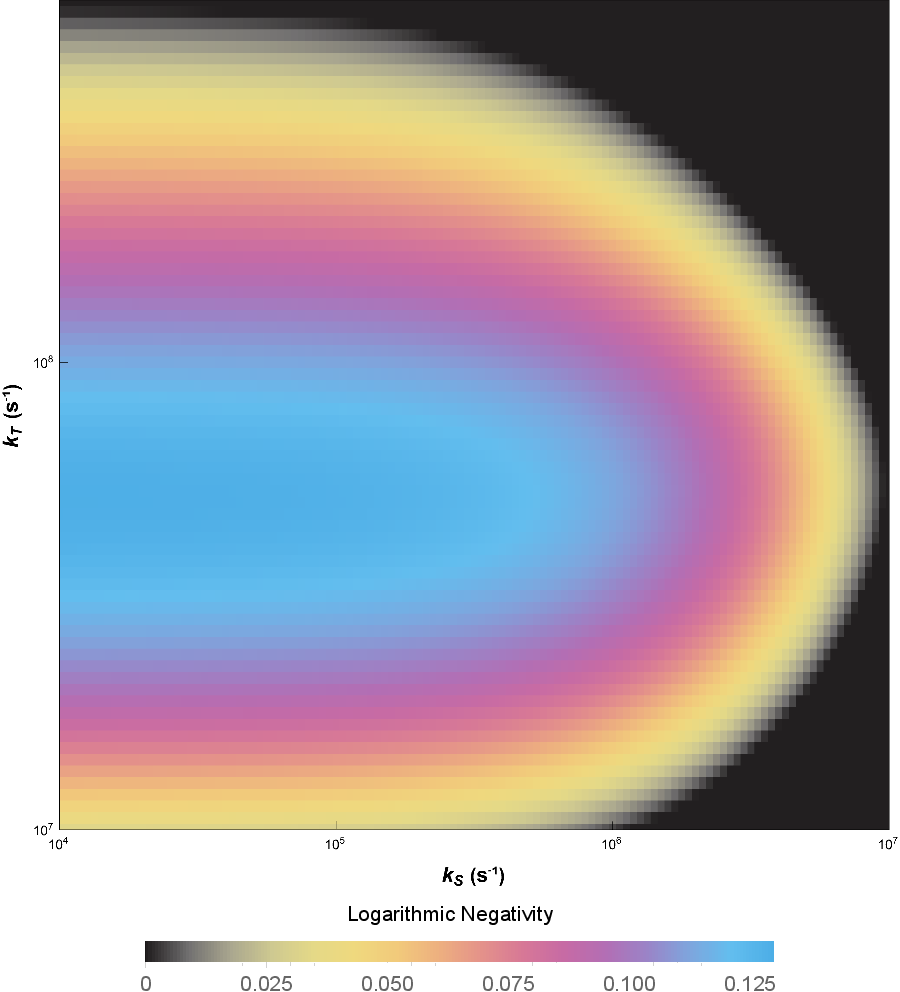}
    \caption{\textit{Nuclear-spin entanglement generation in a five-nucleus model of a photosystem I radical pair.} Nuclear-spin logarithmic negativity after the first singlet-born RP cycle of the five-nucleus $P700^{\bullet+}$--$A_1^{\bullet-}$ model, starting from maximally mixed nuclear spins. The bipartition comprises the two P700 protons and the three $A_1$ protons, with subsystem dimensions $4$ and $8$. The isotropic hyperfine couplings are $0.27$ and $0.22~\mathrm{mT}$ on P700 and $0.35~\mathrm{mT}$ for each $A_1$ proton. Nuclear-spin entanglement is generated over a range of recombination rates, reaching a maximum logarithmic negativity of $0.13$. We set the magnetic field $B$, electron exchange coupling $J$, and dipolar coupling $D$ to zero. The calculation uses $r_A=r_B=10^6~\mathrm{s}^{-1}$ and a logarithmic grid with 0.025-decade spacing along both axes ($121\times72$ points), within $k_S=10^4$--$10^7$ and $k_T=10^7$--$6\times10^8~\mathrm{s}^{-1}$. The evolution over five cycles is shown in Fig.~\ref{fig:S3}.}
    \label{fig:3}
\end{figure}

\textit{Results---} We first examine nuclear-spin entanglement generation in the two-nucleus model. Hyperfine interactions can transfer electronic entanglement to the nuclei, while spin-selective recombination can itself generate electronic entanglement during RP dynamics. To determine when nuclear-spin entanglement can arise, we vary the initial electronic state and allow only one reaction channel at a time (Fig.~\ref{fig:S1}). The survey shows that initial electronic entanglement is neither sufficient nor necessary for nuclear-spin entanglement generation. For example, a singlet initial state with singlet-only recombination produces no nuclear entanglement, whereas the separable electronic state $\ket{T_+}$ can produce nuclear entanglement through the same channel. The full survey is given in the Supplemental Material (Fig.~\ref{fig:S1}).

To illustrate how spin-selective recombination can generate electronic entanglement, consider two electrons initially in the separable state $\ket{01}=(\ket S+\ket{T_0})/\sqrt{2}$. With only triplet recombination, and omitting Hamiltonian evolution and spin relaxation, the unnormalized surviving state evolves as
\begin{equation}
\ket{01}\longrightarrow\frac{\ket S+e^{-k_Tt/2}\ket{T_0}}{\sqrt{2}}.
\label{eq:electronic_entanglement_illustration}
\end{equation}
As the triplet component decays, the normalized surviving state becomes entangled and approaches the singlet state. Hyperfine interactions provide a route for transferring electronic entanglement to the nuclei.

We now focus on a singlet-born RP in which both singlet and triplet recombination channels are present. We describe their combined effect using the Haberkorn reaction operator. Nuclear-spin entanglement is generated predominantly when $k_T$ is greater than, or at least comparable to, $k_S$, with the largest values generally appearing after the first cycle. Its behaviour over successive cycles depends on both the relative and absolute values of these recombination rates. Figure~\ref{fig:1b} classifies this behaviour over ten cycles. When triplet recombination is much faster than singlet recombination, the logarithmic negativity generally alternates between zero and nonzero values; this region also contains the largest first-cycle entanglement. When the recombination rates are comparable, nondecreasing entanglement is observed, with this region extending to $k_S>k_T$. Other nonmonotonic behaviour lies predominantly between this region and the region of alternating zero and nonzero entanglement. Transient entanglement occurs along parts of the boundary with the region where no entanglement is detected. As the cycles progress, the region in which entanglement remains nonzero from one cycle to the next becomes broader (Fig.~\ref{fig:S2}). Figure~\ref{fig:1c} shows representative trajectories from the four regions where entanglement is detected. The classification criteria are given in the Supplemental Material.

We next examine how the magnetic field and hyperfine couplings affect nuclear-spin entanglement over ten cycles at point $\mathrm{I}$. Figure~\ref{fig:2a} compares different magnetic fields for equal hyperfine couplings, $a_A=a_B=1~\mathrm{mT}$. The largest entanglement is obtained at zero field, and increasing the field generally reduces the oscillation amplitudes. This behaviour is physically reasonable, since a sufficiently small Zeeman interaction allows the hyperfine dynamics to dominate the Hamiltonian. Figure~\ref{fig:2b} compares different $a_B$ values at $B=0.05~\mathrm{mT}$, with $a_A=1~\mathrm{mT}$ fixed. Couplings close to $a_A$ produce larger entanglement over successive cycles. The hyperfine interaction couples the electronic and nuclear spin dynamics, and comparable hyperfine frequencies favour efficient transfer of entanglement between the electronic and nuclear spin degrees of freedom.

We next examine whether nuclear-spin entanglement also arises in a more realistic, larger nuclear-spin system using experimentally reported hyperfine couplings from a biochemical RP. We consider a five-nucleus model of $P700^{\bullet+}\!-\!A_1^{\bullet-}$ in photosystem I~\cite{polm1998secondary}. This RP forms in an electronic singlet state following photoexcitation and electron transfer. When further electron transfer beyond $A_1^{\bullet-}$ is blocked, for example by prereducing the downstream iron--sulfur centres~\cite{polm1998secondary}, the RP can recombine through singlet and triplet channels. We retain the nuclei with the largest isotropic HFIs in the experimental parameter sets summarized in Table~\ref{tab:psi_hfi}: two protons on $P700^{\bullet+}$, with couplings of $+0.27$ and $+0.22~\mathrm{mT}$~\cite{chestnut2021electronic}, and three equivalent methyl protons on $A_1^{\bullet-}$, each with a coupling of $+0.35~\mathrm{mT}$~\cite{niklas2009electronic}. All other nuclei are omitted. We set $B=J=D=0$, where $J$ and $D$ denote electron exchange and dipolar couplings, respectively, and include electron-spin relaxation at $10^6~\mathrm{s}^{-1}$ on each radical. The calculated bipartition comprises the two P700 protons and the three $A_1$ protons, with subsystem dimensions $4$ and $8$.

Over five RPM cycles, we find positive logarithmic negativity in cycles 1, 3, and 5. In the first cycle, the entangled region spans singlet recombination rates of roughly $10^4$--$10^7~\mathrm{s}^{-1}$ and triplet recombination rates of roughly $10^7$--$6\times10^8~\mathrm{s}^{-1}$ (Fig.~\ref{fig:3}). The maximum logarithmic negativity is approximately $0.13$, occurring at $k_S=10^4~\mathrm{s}^{-1}$ and $k_T\approx5.0\times10^7~\mathrm{s}^{-1}$. The maxima decrease to $0.014$ and $8.3\times10^{-4}$ in cycles 3 and 5, respectively, while the entangled region becomes smaller with each successive odd cycle (Fig.~\ref{fig:S3}). Cycles 2 and 4 give zero logarithmic negativity throughout the scanned region. In this $4\times8$ partition, positive logarithmic negativity is a sufficient but not necessary condition for entanglement~\cite{vidal2002computable}. Figure~\ref{fig:3} shows the first cycle, and Fig.~\ref{fig:S3} shows all five cycles.

\textit{Discussion---} In this work, we investigated the generation of nuclear-spin entanglement through the radical pair mechanism. Our results show that nuclear spins can indeed become entangled through repeated cycles of radical pair dynamics, and we identify the conditions under which this entanglement is favoured. In particular, the generation and persistence of entanglement depend strongly on the initial electronic state, the reaction channels, the relative singlet and triplet recombination rates, the hyperfine interactions, and the applied magnetic field. We further find that nuclear-spin entanglement can arise in a five-nucleus model incorporating experimentally reported hyperfine couplings from the photosynthetic radical pair P700--$A_1$, suggesting that its generation may be possible in biochemical radical pairs.

An important question is how to detect the nuclear-spin entanglement predicted here experimentally. The reaction yield is experimentally accessible, and its variation over successive cycles generally follows that of the nuclear-spin entanglement in the examples shown in Fig.~\ref{fig:S4}. The yield can, however, continue to vary even when the logarithmic negativity vanishes. Yield variations therefore reflect changes in the nuclear spin state more generally and cannot by themselves definitively establish entanglement. NMR-based quantum state tomography of the nuclear spins might allow unambiguous experimental verification~\cite{roy2010tomography}.

Beyond detecting nuclear-spin entanglement, an important future direction is to investigate its potential applications. For example, entanglement can enhance the precision of quantum sensing protocols~\cite{degen2017quantum}, motivating investigation of whether RPM-generated nuclear-spin entanglement could provide similar advantages. The potential relevance of this entanglement in biological systems is another subject for future investigation.

\textit{Acknowledgement---}
This work involved the use of the Advanced Research Computing (ARC) cluster at the University of Calgary.

This work was supported by the Natural Sciences and Engineering Research Council of Canada through its Discovery Grant (RGPIN-2020-03945) as well as the Alliance Quantum Consortia Grant `Quantum Enhanced Sensing and Imaging' (ALLRP 578468 - 22), and by the National Research Council of Canada through its Quantum Sensing Challenge Program (CSTIP grant QSP 022). 
\vspace{3mm}

\textit{Author contributions---}
CS conceived the project; R performed the theoretical modelling and calculations with help from HZH and CS; R wrote the original draft with feedback from HZH and CS; R, HZH, and CS reviewed and edited the final version.
\vspace{3mm}

\textit{Code availability---}
The Mathematica notebooks and numerical data accompanying the manuscript are available at the following GitHub repository:
\\
\url{https://github.com/RishabhQuantum/RPM-generated-nuclear-spin-entanglement_v1-}
\vspace{3mm}

\textit{AI use statement---}
The authors developed the core analytical argument and original computational code. ChatGPT/Codex (OpenAI, Astra) assisted with further code development, analytical and numerical checks, literature searches, and manuscript and figure preparation. The authors directed this assistance, reviewed the code’s logic and consistency with the model, and reviewed and revised the AI-assisted material. The authors take responsibility for the final manuscript.

\bibliography{references}

\clearpage
\onecolumngrid

\section*{Supplemental Material}

\setcounter{figure}{0}
\setcounter{table}{0}
\renewcommand{\thetable}{S\arabic{table}}
\renewcommand{\thefigure}{S\arabic{figure}}

\subsection*{Initial-state and reaction-channel survey}

\begin{figure}[h!]
    \centering
    \includegraphics[width=0.6\linewidth]{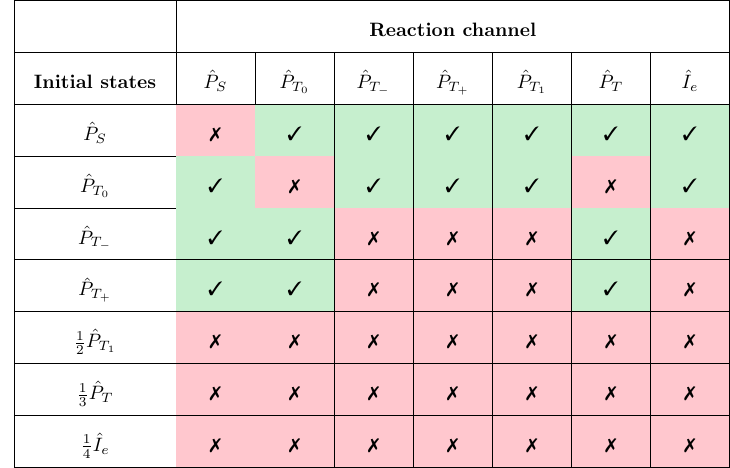}
  
    \caption{\textit{Dependence of nuclear-spin entanglement generation on electronic preparation and reaction channel.} Each row specifies the electronic state prepared at the start of every cycle, and each column shows the sole reaction channel present. The two nuclear spins are initially maximally mixed. A check indicates nuclear-spin logarithmic negativity greater than $10^{-6}$ in at least one of the ten cycles and parameter settings tested; a cross indicates that none was detected within the survey. The results show that neither an entangled initial electronic state nor an entangled reaction projector alone guarantees nuclear-spin entanglement generation: their combination matters. The parameter grid is specified in the accompanying text. Here, $\hat P_{T_1}=\hat P_{T_+}+\hat P_{T_-}$.}
    \label{fig:S1}
\end{figure}

In Fig.~\ref{fig:S1}, only the reaction channel indicated at the top of each column is present. Note that in Figs.~\ref{fig:1}--\ref{fig:3}, both singlet and triplet recombination channels are present simultaneously. At each cycle, the electrons are reset to the state shown in the row. The nuclear spins are initially maximally mixed.

\Needspace{120pt}
Each state--channel combination is tested over ten cycles using $a_A=1\,\mathrm{mT}$, $r_A=r_B=10^6\,\mathrm{s}^{-1}$, and all combinations of
\begin{equation}
\begin{aligned}
a_B&\in\{0.1,0.5,0.9,1,1.1,1.5,10\}\,\mathrm{mT},\\
B&\in\{0.05,0.5,1,5\}\,\mathrm{mT},\\
k&\in\{10^4,10^5,10^6,10^7,10^8\}\,\mathrm{s}^{-1}.
\end{aligned}
\end{equation}
A check indicates logarithmic negativity greater than $10^{-6}$ in at least one tested cycle and parameter setting. A cross indicates none detected within this survey. The estimated numerical error in the logarithmic negativity is of order $10^{-10}$, well below the $10^{-6}$ detection threshold.

This table highlights the role of spin-selective recombination in generating nuclear-spin entanglement. The Hamiltonian $\hat H=\hat H_A\otimes\hat I_B+\hat I_A\otimes\hat H_B$ and local electron relaxation cannot create entanglement across the two radical subsystems from a state separable across that partition. Hyperfine interactions can, however, transfer pre-existing electron-spin entanglement to the nuclear spins. Spin-selective recombination can also generate electronic entanglement from an initially separable state, as illustrated by the triplet-only example in the main text.

\FloatBarrier

\clearpage
\subsection*{Entanglement behaviour classification}
\begin{figure}[h!]
    \centering
    \includegraphics[width=0.9\linewidth]{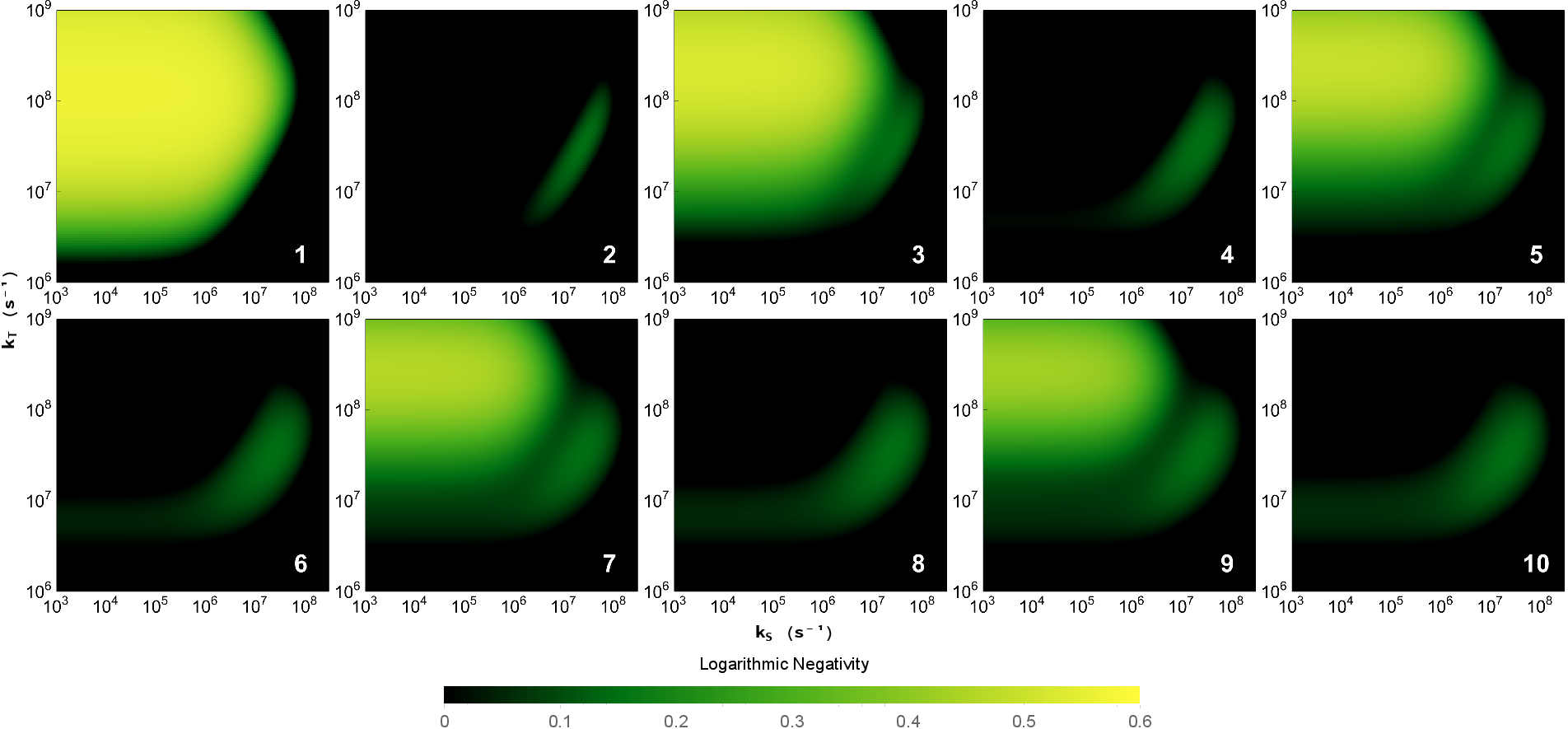}
    \caption{\textit{Evolution of nuclear-spin entanglement over repeated radical pair cycles.} Nuclear-spin logarithmic negativity in the $k_S$--$k_T$ plane after each of the first ten singlet-born RP cycles, starting from maximally mixed nuclear spins. Each radical contains one spin-$1/2$ nucleus. The maps show pronounced differences between odd and even cycles. As the cycles progress, the region of reaction-rate space in which entanglement remains nonzero from one cycle to the next becomes broader. The panels use the same $220\times121$ grid as Fig.~\ref{fig:1b}, with 0.025-decade spacing along both axes, within $k_S=10^3$--$3\times10^8$ and $k_T=10^6$--$10^9~\mathrm{s}^{-1}$. The parameters are $a_A=a_B=1~\mathrm{mT}$, $B=0.05~\mathrm{mT}$, and $r_A=r_B=10^6~\mathrm{s}^{-1}$. All panels share a colour scale from 0 to 0.6.}
    \label{fig:S2}
\end{figure}

Figure~\ref{fig:S2} shows the nuclear-spin logarithmic negativity after each of the first ten cycles. Figure~\ref{fig:1b} uses a logarithmic grid with 0.025-decade spacing along both axes ($220\times121$ points) within $10^3\leq k_S\leq3\times10^8$ and $10^6\leq k_T\leq10^9~\mathrm{s}^{-1}$. The ten-cycle trajectories are classified using a tolerance of $10^{-6}$. The criteria are applied in the following order. If no value exceeds this tolerance, the trajectory is classified as no entanglement detected (grey). If entanglement is detected earlier but is zero within tolerance in the final two cycles, it is classified as transient entanglement (yellow). Trajectories that alternate between zero and nonzero entanglement at every successive cycle are classified as zero/nonzero oscillations (orange). Of the remaining trajectories, those with no decrease greater than $10^{-6}$ between successive cycles are classified as nondecreasing entanglement (green). All remaining trajectories are classified as other nonmonotonic entanglement (blue).

\FloatBarrier

\clearpage
\subsection*{Nuclear-spin entanglement in the five-nucleus model of a photosystem I radical pair}

The five-nucleus model retains the nuclei with the largest isotropic HFIs in the experimental parameter sets in Table~\ref{tab:psi_hfi}. Only the isotropic components are used; the remaining nuclei and hyperfine anisotropy are omitted. Poluektov et al.~\cite{poluektov2005bidirectional} report exchange and dipolar coupling parameters of $J=5~\mu\mathrm{T}$ and $D=-0.17~\mathrm{mT}$ for the A-branch $P700^{\bullet+}A_{1A}^{\bullet-}$ radical pair in \textit{Synechococcus lividus}, a different species from \textit{Thermosynechococcus elongatus} used for the HFI values in Table~\ref{tab:psi_hfi}. We neglect both interactions in our five-nucleus model, setting $J=D=0$.

\begin{table}[h!]
\centering
\caption{\textit{Isotropic hyperfine couplings underlying the five-nucleus PSI model.} Values for $P700^{\bullet+}$ and $A_1^{\bullet-}$ in photosystem I of \textit{Thermosynechococcus elongatus} are given in MHz, from Table~1 of Chestnut et al.~\cite{chestnut2021electronic} and Table~1 of Niklas et al.~\cite{niklas2009electronic}, respectively. The model uses P700 nuclei H1 and H2 ($a_{\mathrm{iso}}=7.7$ and $6.2$ MHz) and the three $A_1$ position-2 methyl protons ($a_{\mathrm{iso}}=+9.8$ MHz each).}
\label{tab:psi_hfi}
\small
\begin{tabular}{lr}
\hline
$P700^{\bullet+}$ nucleus & $a_{\mathrm{iso}}$ (MHz) \\
\hline
H1 & $7.7\pm0.4$ \\
H2 & $6.2\pm0.3$ \\
N1 & $2.8\pm0.4$ \\
N2 & $2.3\pm0.3$ \\
N3 & $1.4\pm0.2$ \\
N4 & $1.2\pm0.3$ \\
N5 & $0.47\pm0.07$ \\
N6 & $0.45\pm0.07$ \\
\hline
\end{tabular}
\par\smallskip
\begin{tabular}{lrr}
\hline
$A_1^{\bullet-}$ nucleus & \multicolumn{2}{c}{$a_{\mathrm{iso}}$ (MHz)} \\
 & expt & DFT \\
\hline
position 2 (CH$_3$) & $+9.8$ & $+10.51$ \\
position 3 (CH$_2$) & & $+0.29,\;+3.12$ \\
position 7 ($\alpha$) & & $-3.19$ \\
position 5 ($\alpha$) & & $-1.82$ \\
position 6 ($\alpha$) & & $-0.89$ \\
position 8 ($\alpha$) & & $+0.46$ \\
H-bond & & $-0.28$ \\
\hline
\end{tabular}

\end{table}
\FloatBarrier

\begin{figure}[h!]
    \centering
    \includegraphics[width=0.9\linewidth]{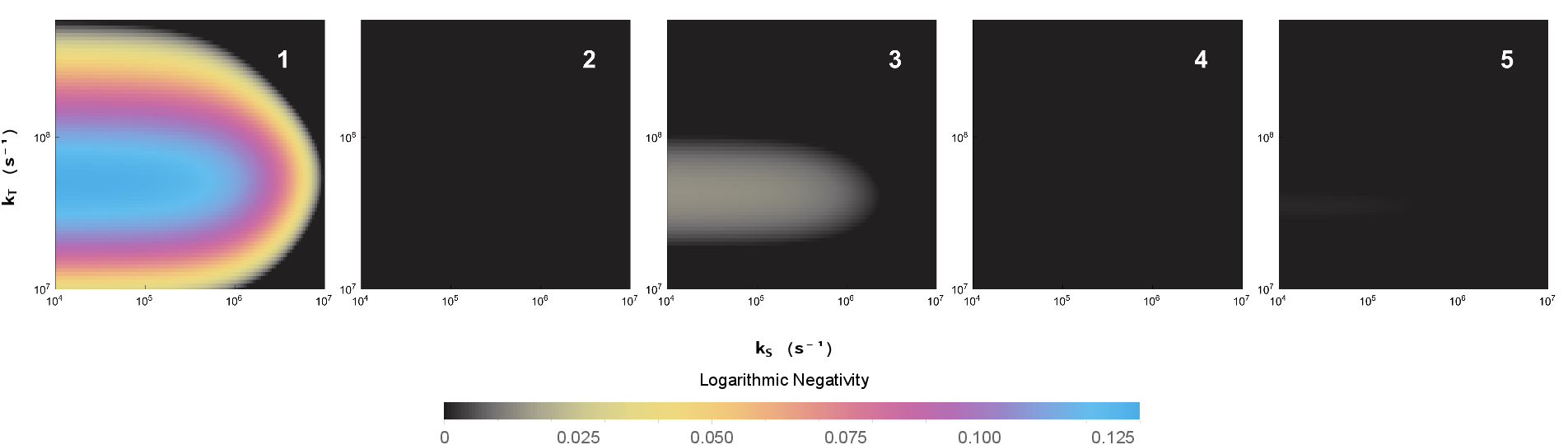}
    \caption{\textit{Cycle-dependent nuclear-spin entanglement in the five-nucleus model of a photosystem I radical pair.} Nuclear-spin logarithmic negativity across the bipartition comprising the two P700 protons and the three $A_1$ protons after each of five singlet-born RP cycles, starting from maximally mixed nuclear spins. Entanglement appears in cycles 1, 3, and 5, with decreasing maxima of approximately $0.13$, $0.014$, and $8.3\times10^{-4}$, respectively. Cycles 2 and 4 have zero logarithmic negativity throughout the scanned grid. The isotropic hyperfine couplings are $0.27$ and $0.22~\mathrm{mT}$ on P700 and $0.35~\mathrm{mT}$ for each $A_1$ proton. We set the magnetic field $B$, electron exchange coupling $J$, and dipolar coupling $D$ to zero, with $r_A=r_B=10^6~\mathrm{s}^{-1}$. Each panel uses the same $121\times72$ grid as Fig.~\ref{fig:3}, with 0.025-decade spacing along both axes, within $k_S=10^4$--$10^7$ and $k_T=10^7$--$6\times10^8~\mathrm{s}^{-1}$. All panels share a colour scale from 0 to 0.13.}
    \label{fig:S3}
\end{figure}

\FloatBarrier

\clearpage

\subsection*{Singlet reaction yield and nuclear spin purity}

\begin{figure}[h!]
    \centering
    \includegraphics[width=0.95\linewidth]{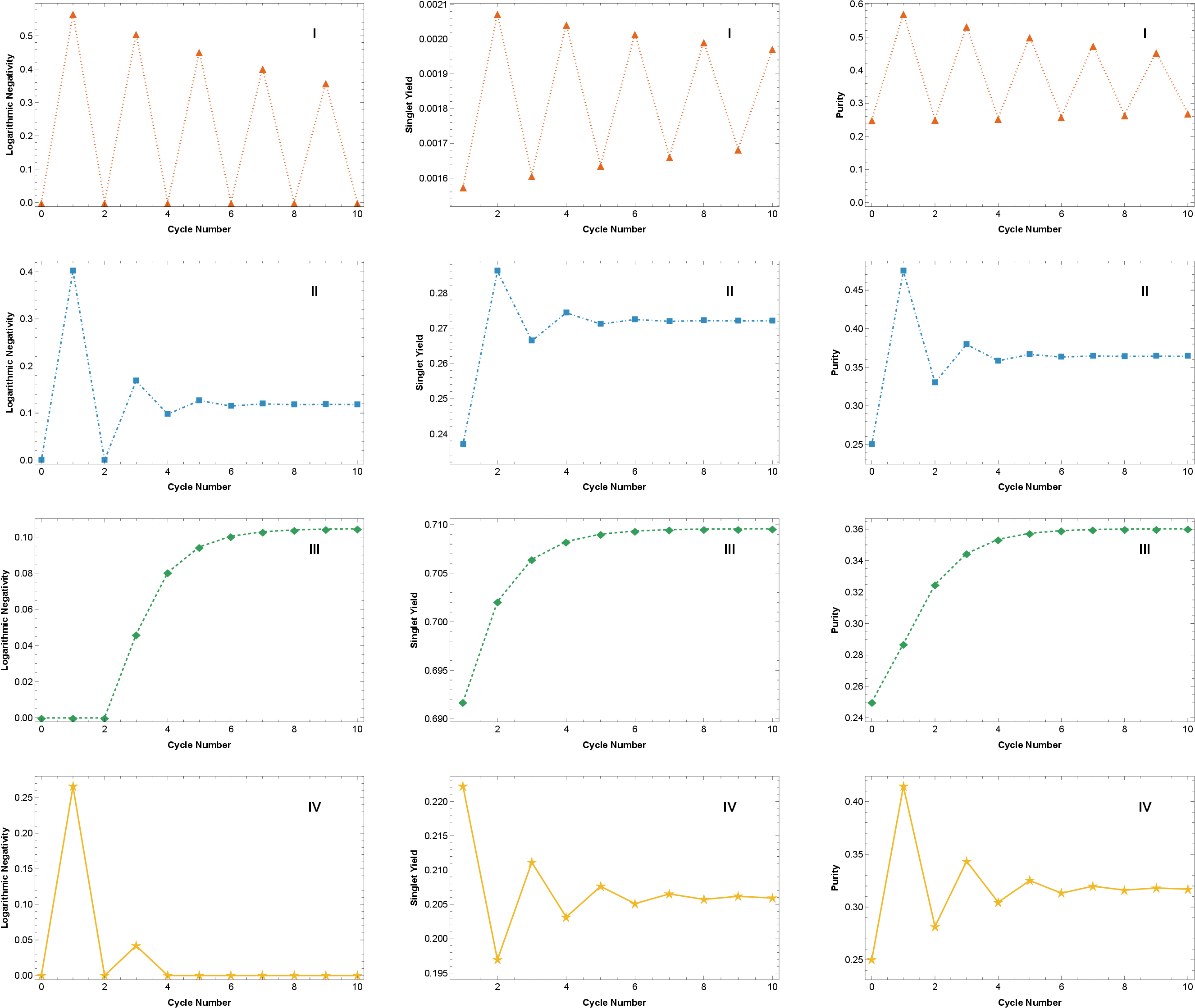}
   
    \caption{\textit{Singlet reaction yield depends on the nuclear spin state, not solely on entanglement.} Nuclear-spin logarithmic negativity (left column), singlet reaction yield (centre column), and nuclear spin purity (right column) over ten singlet-born RP cycles, starting from maximally mixed nuclear spins. Rows 1--4 correspond to points I--IV in Fig.~\ref{fig:1b}, with $(k_S,k_T)=(10^5,10^8)$, $(10^7,4\times10^7)$, $(6\times10^7,3\times10^7)$, and $(2\times10^7,4\times10^8)~\mathrm{s}^{-1}$, respectively. Although the singlet yield varies alongside nuclear-spin entanglement, it continues to vary even when the entanglement vanishes, illustrating that yield variations cannot be attributed solely to entanglement. Entanglement and purity at cycle $m$ describe the output nuclear spin state $\hat\rho_{\mathrm n}^{(m)}$, whereas the yield is generated during that cycle from the input state, $\hat\rho_{\mathrm n}^{(0)}$ for the first cycle and $\hat\rho_{\mathrm n}^{(m-1)}$ for subsequent cycles. Cycle zero denotes the initial nuclear spin state and has no reaction yield. The remaining parameters are $a_A=a_B=1~\mathrm{mT}$, $B=0.05~\mathrm{mT}$, and $r_A=r_B=10^6~\mathrm{s}^{-1}$. Vertical scales vary between rows.}
    \label{fig:S4}
\end{figure}

Figure~\ref{fig:S4} compares the reaction yield, nuclear-spin logarithmic negativity, and nuclear spin purity over repeated cycles at points I--IV in Fig.~\ref{fig:1b}. The yield and logarithmic negativity show similar trends across these examples. The yield continues to vary even during cycles in which the logarithmic negativity remains zero, illustrating why changes in yield alone cannot definitively establish nuclear-spin entanglement.

\FloatBarrier

\end{document}